\documentclass[twocolumn,showpacs]{revtex4}
\usepackage{amsmath,amsthm,amssymb,amscd,float}
\usepackage{graphicx}
\usepackage{lmodern}
\newcommand{\de}{\delta}

\newcommand{\vep}{\varepsilon}

\newcommand{\si}{\sigma}

\newcommand{\vt}{\vartheta}

\newcommand{\De}{\Delta}

\newcommand{\bde}{\boldsymbol{\de}}
\newcommand{\bk}{\mathbf{k}}

\newcommand{\bn}{\mathbf{n}}

\newcommand{\cE}{{\mathcal E}}

\newcommand{\cH}{{\mathcal H}}

\newcommand{\cP}{{\mathcal P}}

\newcommand{\cZ}{{\mathcal Z}}
\def\Esc{E^{\mathrm{sc}}}
\def\Hsc{H^{\mathrm{sc}}}
\def\Zsc{Z^{\mathrm{sc}}}
\newcommand{\pa}{\partial}

\let\ds\displaystyle
\let\ni\noindent
\newcommand{\ms}{\mspace{1mu}}
\renewcommand{\le}{\leqslant}
\renewcommand{\ge}{\geqslant}

\renewcommand{\geq}{\geqslant}
\newcommand{\tr}{\operatorname{tr}}

\renewcommand{\mod}{\operatorname{mod}}

\newcommand{\e}{\mathrm{e}}

\newcommand{\diff}{\mathrm{d}}
\newcommand{\smax}{s_{\mathrm{max}}}
\newcounter{ex}
\def\cond{\stepcounter{ex}\hskip-.75cm
\makebox[.55cm][r]{(\roman{ex})}\hskip.2cm}
\begin{document}
\title{The Berry--Tabor conjecture for spin chains of Haldane--Shastry type}
\author{J.C. \surname{Barba}}%
\email{jcbarba@fis.ucm.es} \author{F. \surname{Finkel}}%
\email{ffinkel@fis.ucm.es} \author{A. \surname{Gonz\'{a}lez-L\'{o}pez}}
\email{artemio@fis.ucm.es}
\author{M.A. \surname{Rodr\'{\i}guez}}%
\email{rodrigue@fis.ucm.es} \affiliation{Departamento de F\'{\i}sica
  Te\'{o}rica II, Universidad Complutense, 28040 Madrid, Spain}
\date{April 23, 2008}
\begin{abstract}
  According to a long-standing conjecture of Berry and Tabor, the
  distribution of the spacings between consecutive levels of a ``generic''
  integrable model should follow Poisson's law. In contrast, the spacings
  distribution of chaotic systems typically follows Wigner's law. An important
  exception to the Berry--Tabor conjecture is the integrable spin chain with
  long-range interactions introduced by Haldane and Shastry in 1988, whose
  spacings distribution is neither Poissonian nor of Wigner's type. In this
  letter we argue that the cumulative spacings distribution of this chain
  should follow the ``square root of a logarithm'' law recently proposed by us
  as a characteristic feature of all spin chains of Haldane--Shastry type. We
  also show in detail that the latter law is valid for the rational
  counterpart of the Haldane--Shastry chain introduced by Polychronakos. 
\end{abstract}
\maketitle
\section{Introduction}

It is well known~\cite{GMW98} that the distribution of (suitably normalized) spacings between consecutive
eigenvalues for the Gaussian ensembles in random matrix theory is approximately given by
Wigner's surmise
$$
p(s)=(\pi/2)\ms s\exp(-\pi s^2/4)\,.
$$
This behavior, which was first observed in the spectra of complex atomic
nuclei, also seems
to be characteristic of quantum chaotic systems like polygonal billiards~\cite{Me04}.
On the other hand, for a ``generic'' quantum integrable system
Berry and Tabor have conjectured~\cite{BT77} that the spacings distribution
should instead follow Poisson's law $p(s)=\e^{-s}$.
Over the years, this conjecture has been verified for many
integrable models of physical interest, such as the Heisenberg chain, the
$t$-$J$ model, the Hubbard model~\cite{PZBMM93} and the chiral Potts
model~\cite{AMV02}. In a recent paper~\cite{FG05} it has been shown that there is
an important exception to Berry and Tabor's conjecture, namely the
integrable spin chain introduced by Haldane and Shastry in 1988~\cite{Ha88,Sh88}, whose spacings distribution follows neither Poisson's nor Wigner's law. It is the purpose of this letter to
gain a deeper understanding of this fact, and to explore whether this property is
shared by the analogous chain introduced by Polychronakos in ref.~\cite{Po93}.

The (antiferromagnetic) Haldane--Shastry (HS) spin chain describes a system of
$N$ spins equally spaced on a circle with long-range exchange interactions inversely proportional
to the square of the chord distance between the spins. More precisely, its Hamiltonian
is given by
\begin{equation}\label{HS}
\cH_{\mathrm{HS}}=\frac12\sum_{i<j}\sin(\vt_i-\vt_j)^{-2}(1+S_{ij}),\qquad \vt_k=\frac{k\pi}N\,,
\end{equation}
where $S_{ij}$ is the operator permuting the $i$-th and $j$-th spins.
(Unless otherwise stated, throughout the paper all sums and products run from $1$ to $N$.)
The HS chain is closely related to the Hubbard model; for instance, the chain's ground state
coincides exactly with Gutzwiller's variational wave function for the Hubbard model when the
on-site interaction tends to infinity~\cite{Hu63,Gu63,GV87}. Another important characteristic
of the HS chain is its connection with the Sutherland spin model of $A_N$
type~\cite{Su71,Su72,HH92,HW93,MP93}, from which it can be obtained by means of the so-called ``freezing trick''~\cite{Po93}. This technique essentially consists in taking the strong coupling limit
in the spin Sutherland model, so that the internal and dynamical degrees of freedom decouple and the
spins become frozen at the equilibrium positions of the scalar part of the potential, which are
precisely the chain sites $\vt_k$. This procedure can in fact be applied to all spin models of Calogero and
Sutherland type, associated with both the $A_N$ and $BC_N$ roots systems in Olshanetsky and Perelomov's
scheme~\cite{OP83}. For instance, the spin Calogero model of $A_N$ type~\cite{Ca71,MP93}
\begin{equation}\label{H}
H=-\sum_i\pa_{x_i}^2+a^2\sum_i x_i^2+a\sum_{i\neq j}\frac{a+S_{ij}}{(x_i-x_j)^2}
\end{equation}
yields the Hamiltonian of the so-called Polychronakos--Frahm (PF) spin chain~\cite{Po93,Fr93}
\begin{equation}\label{cH}
\cH_{\mathrm{PF}}=\sum_{i<j}(\xi_i-\xi_j)^{-2}(1+S_{ij})\,.
\end{equation}
The sites of this chain are the coordinates of the unique critical point of the scalar part of the potential
of the Hamiltonian~\eqref{H} in the domain $x_1<\cdots<x_N$, namely the numbers $\xi_1<\cdots<\xi_N$
determined by the algebraic system
\begin{equation}
\xi_i=2\sum_{j,j\ne i}\frac1{(\xi_i-\xi_j)^3}\,,\qquad i=1,\dots,N\,.
\end{equation}
Remarkably, these numbers are just the roots of the Hermite polynomial of degree $N$, as first pointed out by Calogero~\cite{Ca77}.

In order to compute the spacings distributions of a spectrum,
it is first necessary to transform the ``raw'' spectrum
by applying the so-called \emph{unfolding mapping}~\cite{Ha01}. This
mapping is defined by decomposing the cumulative level density $F(\cE)$ as the
sum of a fluctuating part $F_{\mathrm{f{}l}}(\cE)$ and a continuous part
$\eta(\cE)$, which is then used to transform each energy $\cE_i$,
$i=1,\dots,n$, into an unfolded energy $\eta_i=\eta(\cE_i)$. In this way one
obtains a uniformly distributed spectrum $\{\eta_i\}_{i=1}^n$, regardless of
the initial level density. One finally considers the normalized spacings
$s_i=(\eta_{i+1}-\eta_i)/\De$, where $\De=(\eta_{n}-\eta_1)/(n-1)$ is the mean
spacing of the unfolded energies, so that $\{s_i\}_{i=1}^{n-1}$ has unit mean.
The analysis of the spacings distribution has been carried out so far for the original ($A_N$-type)
HS chain~\cite{FG05}, its  supersymmetric version~\cite{BB06} and, very recently~\cite{BFGR08}, for the
PF chain of $BC_N$-type introduced by Yamamoto and Tsuchiya~\cite{YT96}. The corresponding
spacings distributions were found to be qualitatively very similar, and
differed essentially from both Wigner's and Poisson's distributions.

For the PF chain of $BC_N$ type, we showed in ref.~\cite{BFGR08} that for large $N$ the cumulative spacings
distribution $P(s)\equiv\int_0^s p(t)\ms\diff t$ was approximately given by
\begin{equation}\label{P}
P(s)\simeq 1-\frac{2}{\sqrt\pi\,\smax}\,\sqrt{\log\Big(\frac{\smax}s\Big)}\,,
\end{equation}
where $\smax$ is the maximum spacing. In fact, the above approximation holds for
any spectrum $\cE_1\equiv\cE_{\mathrm{min}}<\cdots<\cE_n\equiv\cE_{\mathrm{max}}$
obeying the following conditions:

{\leftskip.75cm\parindent=0pt\setcounter{ex}{0}%
\cond The energies are equispaced, \emph{i.e.}, $\cE_{i+1}-\cE_i=d$ for $i=1,\dots,n-1$.

\cond The level density (normalized to unity) is approximately given by the Gaussian law
\begin{equation}\label{Gaussian}
g(\cE)=\frac{1}{\sqrt{2\pi}\si}\,\e^{-\frac{(\cE-\mu)^2}{2\si^2}}\,,
\end{equation}
where $\mu$ and $\si$ are respectively the mean and the standard deviation of the spectrum.

\cond $\cE_{\mathrm{max}}-\mu\,,\,\mu-\cE_{\mathrm{min}}\gg\si$.

\cond $\cE_{\mathrm{min}}$ and $\cE_{\mathrm{max}}$ are approximately symmetric with respect to $\mu$, namely
$\vert\cE_{\mathrm{min}}+\cE_{\mathrm{max}}-2\mu\vert\ll\cE_{\mathrm{max}}-\cE_{\mathrm{min}}$.\smallskip

}
\ni As shown in ref.~\cite{BFGR08}, the above assumptions lead to the formula~\eqref{P} with the following explicit
expression for $\smax$:
\begin{equation}\label{smax}
\smax=\frac{(n-1)d}{\sqrt{2\pi}\si}\,.
\end{equation}
It is relatively straightforward to check~\cite{BFGR08} that conditions (i)--(iv) are indeed satisfied by the PF chain
of $BC_N$ type (with $d=1$). In the rest of this letter we shall discuss the applicability of
the approximation~\eqref{P} to the original ($A_N$-type) PF and HS chains. We shall see that conditions (i)--(iv)
are all satisfied by the PF chain, so that eq.~\eqref{P} is guaranteed to work in this case.
On the other hand, we shall explain why the latter formula is still an excellent approximation for the HS chain,
even if not all of the above conditions hold for this chain.

\section{The Polychronakos--Frahm chain}

The initial step in our analysis of the statistical properties of the spectrum
of the PF chain is the explicit knowledge of the partition function, which makes it possible
to compute the energy levels and their degeneracies for relatively large values of $N$.
As first shown by Polychronakos~\cite{Po94}, the partition function can be evaluated in closed form
by applying the freezing trick to the spin Calogero model~\eqref{H}. We shall present here an alternative
expression for the partition function of the PF chain, which is more amenable to numerical computations
than Polychronakos's original formula.

The starting point in our derivation is the identity
$$
H=\Hsc+2a\cH_{\mathrm{PF}}\big|_{\xi_k\to x_k}\,,
$$
where $\Hsc=H\big|_{S_{ij\to 1}}$ is the Hamiltonian of the scalar Calogero model.
As explained in~\cite{BFGR08}, this identity implies
that for large $a$ the energies of $H$ are approximately of the form
$$
E_{ij}\simeq\Esc_i+2a\cE_j\,,
$$
where $\Esc_i$ and $\cE_j$ are two arbitrary eigenvalues of $\Hsc$ and $\cH_{\mathrm{PF}}$.
Although this relation cannot be used directly to compute the chain energies $\cE_j$,
one can use it to express the partition function $\cZ_{\mathrm{PF}}$ of the PF chain
as
\begin{equation}\label{ZZZ}
\cZ_{\mathrm{PF}}(T)=\lim_{a\to\infty}\frac{Z(2aT)}{\Zsc(2aT)}
\end{equation}
in terms of the partition functions $Z$ and $\Zsc$ of $H$ and $\Hsc$, respectively.
Both of these partition functions can be readily computed. Indeed, in the scalar case
the energies are given by~\cite{Ca71}
\begin{equation}\label{Esc}
\Esc_\bn=E_0+2a\sum_i n_i,\quad E_0\equiv aN\big(a(N-1)+1\big),
\end{equation}
with $\bn=(n_1,\dots,n_N)$, where $n_1\ge\cdots\ge n_N$ are nonnegative integers.
Setting $q=\e^{-1/(k_{\mathrm B}T)}$ and neglecting the ground state energy $E_0$
(which will also be subtracted from the energies of $H$), we obtain
$$
\Zsc(2aT)=\sum_{n_1\geq\cdots\geq n_N\geq 0}q^{\sum\limits_i n_i}\,.
$$
Calling $p_j=n_j-n_{j+1}$, $j=1,\dots,N-1$, and $p_N=n_N$, so that
$\sum\limits_in_i=\sum\limits_i\sum\limits_{j\ge i}p_j=\sum\limits_j jp_j$, we finally have
\begin{equation}\label{Zsc}
\Zsc(2aT)=\sum_{p_1,\dots,p_N\ge 0}q^{\sum\limits_j jp_j}=\prod_i\big(1-q^i\big)^{-1}\,.
\end{equation}
The energies of the spin Hamiltonian $H$ are still given by the RHS of~\eqref{Esc}, but now there
is a degeneracy factor $d_\bn$ due to the spin~\cite{BUW99}. More precisely, if $m$ is
the number of internal degrees of freedom ($su(m)$ spin) and
$$
\bn=\big(\overbrace{\vphantom{1}\nu_1,\dots,\nu_1}^{k_1},\dots,
\overbrace{\vphantom{1}\nu_r,\dots,\nu_r}^{k_r}\big),\qquad \nu_1>\cdots>\nu_r\geq0,
$$
then
$$
d_\bn=\prod_{i=1}^r\binom{m}{k_i}\equiv d(\bk)\,,\qquad
\bk=(k_1,\dots,k_r)\,.
$$
Therefore
\begin{equation}\label{Z1}
Z(2aT)=\sum_{\bk\in\cP_N}\sum_{\nu_1>\cdots>\nu_r\ge0}d(\bk)\,q^{\,\sum\limits_{i=1}^rk_i\nu_i}\,,
\end{equation}
where $\cP_N$ denotes the set of ordered partitions of $N$. Calling again
$p_j=\nu_j-\nu_{j+1}>0$, $j=1,\dots,r-1$, and $p_r=\nu_r\ge0$, we have
$$
\sum_{i=1}^rk_i\nu_i=\sum_{i=1}^rk_i\sum_{j=i}^rp_j=\sum_{j=1}^rK_jp_j\,,
$$
where $K_j=\sum\limits_{i=1}^jk_i$. Substituting the previous equation into~\eqref{Z1}
and proceeding as before we easily obtain
\begin{equation}\label{Z}
  Z(2aT)=\sum_{\bk\in\cP_N}
  q^{\sum\limits_{i=1}^{r-1}K_i}\prod_{i=1}^r\frac{\binom{m}{k_i}}
{1-q^{K_i}}\,.
\end{equation}
Since $1\le K_1<\cdots<K_{r-1}<K_r=N$, we can define $N-r$ natural numbers $K'_i$ by
$$
\{K'_1,\dots,K'_{N-r}\}=\{1,\dots,N-1\}-\{K_1,\dots,K_{r-1}\}\,.
$$
{}From eqs.~\eqref{ZZZ},~\eqref{Zsc} and~\eqref{Z}, it immediately follows that the partition
function of the PF chain can be written as
\begin{equation}\label{ZPF}
\cZ_{\mathrm{PF}}(T)=\sum_{\bk\in\cP_N}\prod_{i=1}^r\binom{m}{k_i}\cdot
q^{\sum\limits_{i=1}^{r-1}K_i}\cdot\prod_{i=1}^{N-r}\big(1-q^{K'_i}\big)\,.
\end{equation}
One can show that the latter expression is equivalent to Polychronakos's by arguing
as in ref.~\cite{BBHS07}. Moreover, in the latter reference it is shown that
eq.~\eqref{ZPF} implies that the energies of the PF chain are given by
\begin{equation}\label{cEde}
\cE(\bde)=\sum_{i=1}^{N-1}\de_ii\,,
\end{equation}
where the \emph{motif} $\bde=(\de_1,\dots,\de_{N-1})$ is a sequence of $0$'s and $1$'s with at most
$m-1$ consecutive $0$'s. {}From the previous formula it follows that the
spectrum of the PF chain is a set of consecutive integers, so that condition (i) in the previous section
is satisfied with spacing $d=1$. Hence the number of distinct energies $n$ is simply the difference
$\cE_{\mathrm{max}}-\cE_{\mathrm{min}}+1$, where the minimum and maximum energies can be computed
with the help of eq.~\eqref{cEde}. Indeed, the maximum energy is obviously obtained from the motif $(1,\dots,1)$,
so that
\begin{equation}\label{cEmax}
\cE_{\mathrm{max}}=\sum_{i=1}^{N-1}i=\frac12\,N(N-1)\,.
\end{equation}
On the other hand, the minimum energy corresponds to the motif
\begin{equation}\label{demin}
\bde=\big(\dots,1,\overbrace{0,\dots,0}^{m-1},1,
\overbrace{0,\dots,0}^{m-1}\big),
\end{equation}
in which $\de_i=1$ for $i=N-j\ms m$ with $j=1,\dots,k\equiv\lfloor N/m\rfloor$.
Thus
\begin{align}\label{cEmin}
\cE_{\mathrm{min}}&=\sum_{j=1}^{k}(N-j\ms m)=kN-\frac m2\,k(k+1)\notag\\
&=\frac{N^2}{2m}-\frac N2+\frac{l(m-l)}{2m}\,,\qquad l\equiv N\mod m\,.
\end{align}

\begin{figure}[h]
\centering
\includegraphics[width=8cm]{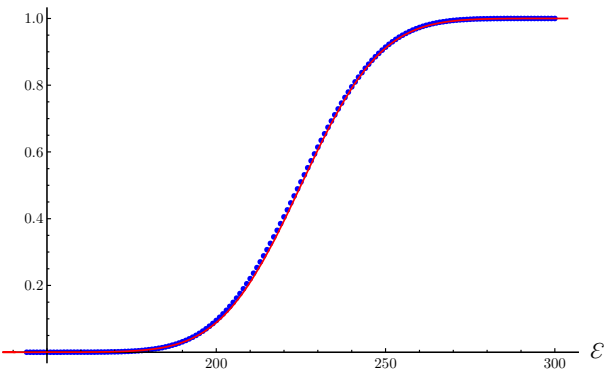}
\caption{Cumulative energy density (at its
discontinuity points) and its continuous approximation
$\int_{-\infty}^\cE g(t)\diff t$ (continuous red line) for the PF chain with
$N=25$ and $m=2$.}
\label{fig:Gaussian}
\end{figure}

\begin{figure}[h]
\centering
\includegraphics[width=8cm]{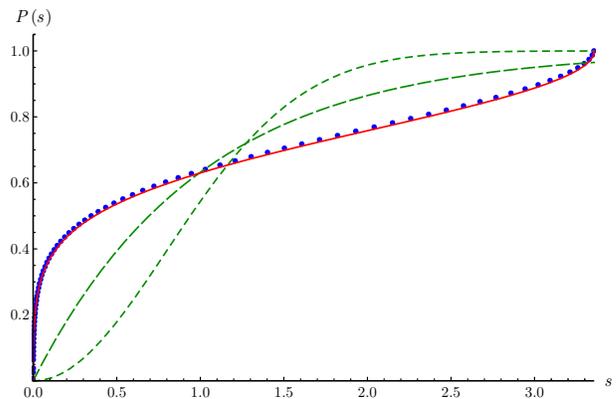}
\caption{Cumulative spacings distribution $P(s)$ and its analytic approximation~\eqref{P}
(continuous red line) for the PF chain with $N=25$ and $m=2$. For convenience, we
have also plotted Poisson's (green, long dashes) and Wigner's (green,
short dashes) cumulative distributions.
\label{fig:spacings-PF}}
\end{figure}

We shall next examine the validity of the second condition in the previous section for the PF chain.
In fact, eq.~\eqref{ZPF} for the partition function turns out to be very efficient for computing
the spectrum of this chain for large values of $N$ (up to $N=26$ for $m=2$ or $N=20$ for $m=3$,
using \textsc{Mathematica}\texttrademark{} on a personal computer). In this way we have ascertained that
the level density obeys the Gaussian law~\eqref{Gaussian}; \emph{cf.}~fig.~\ref{fig:Gaussian} for the case $N=25$
and $m=2$.
The mean energy $\mu$ and the variance $\si$ in eq.~\eqref{Gaussian} can be computed in closed form
for arbitrary values of $N$ and $m$ using several classical identities for the zeros of Hermite polynomials.
More precisely, from the equality $\tr S_{ij}=m^{N-1}$ we easily obtain
\begin{align}
\mu=\frac{\tr\cH_{\mathrm{PF}}}{m^N}
&=\frac12\,\Big(1+\frac1m\Big)\sum_{i\ne j}(\xi_i-\xi_j)^{-2}\notag\\
&=\frac16\,\Big(1+\frac1m\Big)\Big[2N(N-1)-\sum_i\xi_i^2\Big]\,,
\end{align}
where in the last equality we have used eq.~(3.3b) of ref.~\cite{ABCOP79}.
In order to evaluate the last sum, we use the
well-known identity~\cite[eq.~(3.3a)]{ABCOP79}
\begin{equation}\label{3.3a}
\sum_{j,j\ne i}\frac1{\xi_i-\xi_j}=\xi_i\,.
\end{equation}
Since, by antisymmetry,
$$
\sum_{i\ne j}\frac{\xi_i}{\xi_i-\xi_j}=\frac12\,N(N-1)\,,
$$
multiplying eq.~\eqref{3.3a} by $\xi_i$ and summing over $i$ we obtain
\begin{equation}\label{xi2}
\sum_i\xi_i^2=\frac12\,N(N-1)\,,
\end{equation}
and therefore
\begin{equation}\label{mu}
\mu=\frac14\,\Big(1+\frac1m\Big)N(N-1)\,.
\end{equation}
Similarly, using the identity
$$
\tr(S_{ij}S_{kl})=m^{N-2+2\de_{ik}\de_{jl}+2\de_{il}\de_{jk}}\,,
$$
and proceeding as in ref.~\cite{EFGR05}, we obtain
\begin{multline}\label{si20}
\si^2=\frac{\tr(\cH_{\mathrm{PF}}^2)}{m^N}-\frac{(\tr\cH_{\mathrm{PF}})^2}{m^{2N}}\\
=\frac12\Big(1-\frac1{m^2}\Big)\sum_{i\ne j}(\xi_i-\xi_j)^{-4}.
\end{multline}
{}From~\cite[eq.~(3.3d)]{ABCOP79} and~\eqref{xi2} it follows that
\begin{equation}\label{xim4}
45\sum_{i\ne j}(\xi_i-\xi_j)^{-4}=N(N-1)(2N+7)+\sum_i\xi_i^4\,.
\end{equation}
Multiplying eq.~\eqref{3.3a} by $\xi_i^3$ and summing over $i$ we obtain
\begin{align}
\sum_i\xi_i^4&=\sum_{i\ne j}\frac{\xi_i^3}{\xi_i-\xi_j}=
\frac12\sum_{i\ne j}\big(\xi_i^2+\xi_i\xi_j+\xi_j^2\big)\notag\\
&=\Big(N-\frac32\Big)\sum_i\xi_i^2+\frac12\Big(\sum_i\xi_i\Big)^2\notag\\
&=\frac12\,N(N-1)\Big(N-\frac32\Big)\,,\label{xi4}
\end{align}
where we have used eq.~\eqref{xi2} and the obvious identity $\sum_i\xi_i=0$.
From eqs.~\eqref{si20}, \eqref{xim4} and~\eqref{xi4} we finally have
\begin{equation}\label{si2}
\si^2=\frac1{36}\Big(1-\frac1{m^2}\Big)N(N-1)\Big(N+\frac52\Big).
\end{equation}
Equations~\eqref{cEmax}, \eqref{cEmin}, \eqref{mu} and~\eqref{si2} imply that both
$(\cE_{\mathrm{max}}-\mu)/\si$ and $(\mu-\cE_{\mathrm{min}})/\si$ grow as $N^{1/2}$ when $N\to\infty$, so that
condition~(iii) in the previous section is satisfied.
The last of these conditions is also satisfied for large $N$, since by the equations
just quoted $\vert\cE_{\mathrm{min}}+\cE_{\mathrm{max}}-2\mu\vert=O(N)$
while $\cE_{\mathrm{max}}-\cE_{\mathrm{min}}=O(N^2)$. By the discussion in the previous section,
it follows that the cumulative spacings distribution of the PF chain is approximately given by eq.~\eqref{P}
for large $N$. We have indeed verified that~\eqref{P} holds with great accuracy
by computing $P(s)$ for a wide range of values of $N$ and $m$ using eq.~\eqref{ZPF} for
the partition function. For instance, in the case $N=25$ and $m=2$ presented in fig.~\ref{fig:spacings-PF}
the RHS of~\eqref{P} fits the numerical data with a mean square error of $8.4\times 10^{-5}$.

It should be noted that the parameter $\smax$ in the formula~\eqref{P} for $P(s)$ can be computed explicitly as a function
of $N$ and $m$ using~\eqref{smax} (with $d=1$),~\eqref{si2}, the identity
$n-1=\cE_{\mathrm{max}}-\cE_{\mathrm{min}}$, and eqs.~\eqref{cEmax} and~\eqref{cEmin}.
In particular, for large $N$ the maximum spacing $\smax$ behaves as
$$
\smax=\frac3{\sqrt{2\pi}}\,\sqrt{\frac{m-1}{m+1}}\:N^{1/2}+O\big(N^{-1/2}\big)\,,
$$
just as for the PF chain of $BC_N$ type~\cite{BFGR08}. We thus conclude that for $N\gg 1$ the
spacings distributions of the PF chains of $A_N$ and $BC_N$ types are asymptotically
equal, in spite of the fact that the spectra of these models are quite different.

\section{The Haldane--Shastry chain}

\begin{figure}[h]
\centering
\includegraphics[width=8cm]{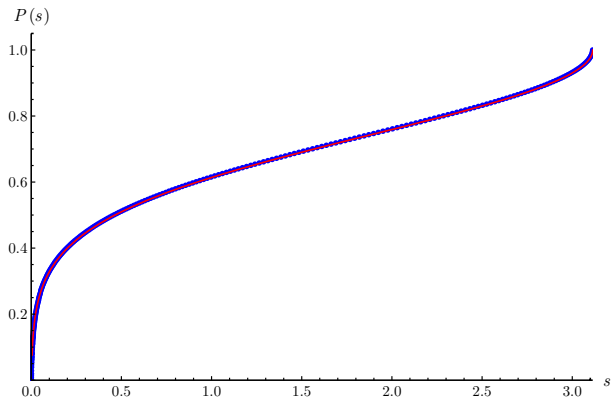}
\caption{Cumulative spacings distribution $P(s)$ and its analytic approximation~\eqref{P}
(continuous red line) for the HS chain with $N=25$ and $m=2$.
The mean square error of the fit is $1.8\times 10^{-5}$.
\label{fig:spacings-HS}}
\end{figure}

It is well-known that the spectrum of the original HS chain~\eqref{HS} is not equispaced, so that
the first condition in section~1 does not hold in this case. However, it was recently
noted~\cite{BFGR08} that the cumulative spacings distribution of this chain
can be still approximated with great accuracy
by a function of the form~\eqref{P}. The parameter $\smax$ is given again by eq.~\eqref{smax}, but now
$d$ is the spacing $\cE_{i+1}-\cE_i$ with highest frequency and $n$ is the total number of energy levels.
As an example, in fig.~\ref{fig:spacings-HS} we present the plot of $P(s)$ and its approximation~\eqref{P}
in the case $N=25$ and $m=2$, for which $d=2$, $n=562$, and hence $\smax\simeq 3.113$.

In the rest of this section we shall analyze why eq.~\eqref{P} provides an excellent approximation to the
cumulative spacings distribution of the HS chain, in spite of the fact that condition (i) in section~1
is not satisfied. We shall start by verifying that the remaining conditions (ii)--(iv) are satisfied
in this case. In the first place, the fact that the level density is approximately Gaussian was established
in ref.~\cite{FG05}, where it was also shown that the mean energy and its standard deviation are
respectively given by
\begin{align}
\mu&=\frac1{12}\,\Big(1+\frac1m\Big)N(N^2-1)\,,\label{muHS}\\[1mm]
\si^2&=\frac1{360}\Big(1-\frac1{m^2}\Big)\,N(N^2-1)(N^2+11)\label{siHS}\,.
\end{align}
As to the third condition, recall that the maximum energy is given
by~\cite{FG05}
\begin{equation}
\cE_\mathrm{max}=\frac N6(N^2-1)\,.
\end{equation}
The minimum energy can be computed from the analogue of eq.~\eqref{cEde}, which for the HS
chain reads~\cite{HHTBP92,FG05}
$$
\cE(\bde)=\sum_{i=1}^{N-1}\de_ii(N-i)\,,
$$
where again $\bde$ is a motif with at most $m-1$ consecutive $0$'s. It can be
shown that the motif
yielding the minimum energy is again given by~\eqref{demin} (or, alternatively, by
the complementary motif $\bde'$ with $\de'_i=\de_{N-i}$), so that
$$
\cE_\mathrm{min}=\sum_{j=1}^k jm(N-jm)\,,
$$
with $k=\lfloor N/m\rfloor$. Evaluating the latter sum we easily obtain
\begin{equation}\label{cEminHS}
\cE_\mathrm{min}=\frac1{6m}\,(N-l)(N+m-l)(N-m+2l)\,,
\end{equation}
where $l\equiv N\mod m$. Equations~\eqref{muHS}--\eqref{cEminHS} imply that (as for the PF chain)
$(\cE_{\mathrm{max}}-\mu)/\si$ and $(\mu-\cE_{\mathrm{min}})/\si$ grow as $N^{1/2}$ when $N\to\infty$, so that
condition~(iii) is clearly satisfied. Finally, from the previous equations
it follows that $\vert\cE_{\mathrm{min}}+\cE_{\mathrm{max}}-2\mu\vert=O(N)$
while $\cE_{\mathrm{max}}-\cE_{\mathrm{min}}=O(N^3)$, which implies condition (iv).

\begin{figure}[h]
\centering
\includegraphics[width=8cm]{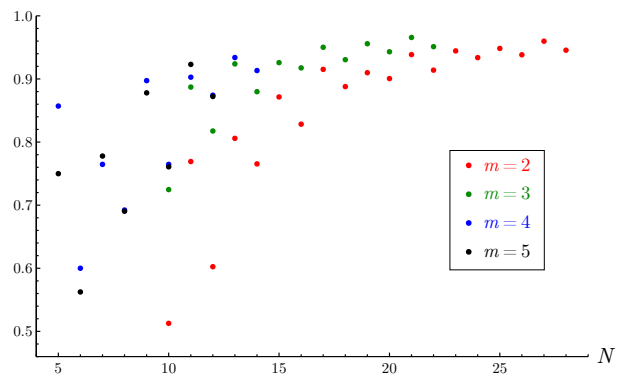}
\caption{Ratio $\ds\frac{\#(d_i=d)}{n-1}$ for the HS chain with $m=2,3,4,5$ and $N$ up to $28$.
\label{fig:percent}}
\end{figure}

\begin{figure}[h]
\centering
\includegraphics[width=8cm]{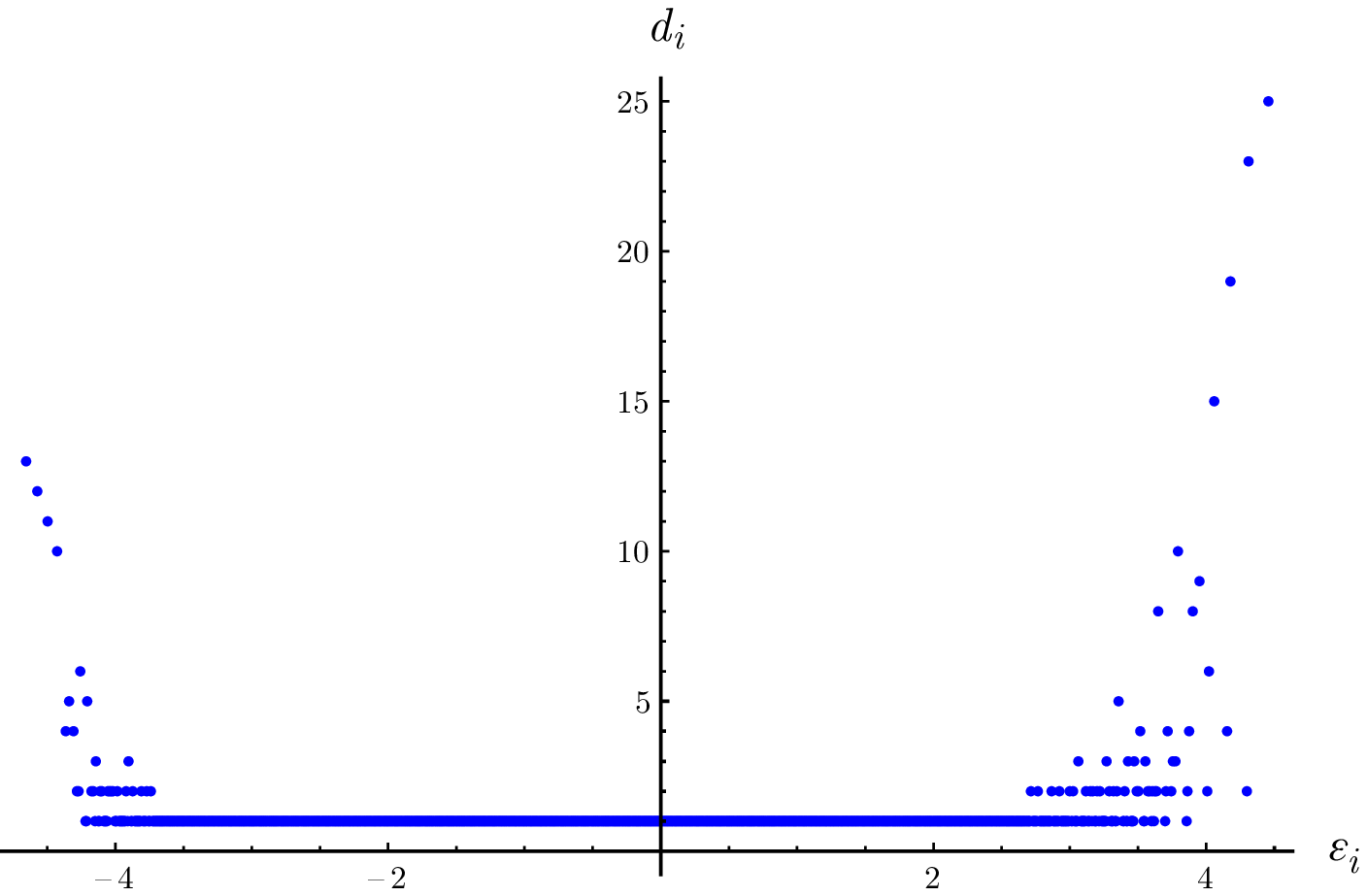}
\caption{Differences $d_i\equiv\cE_{i+1}-\cE_i$ versus $\vep_i\equiv(\cE_i-\mu)/\si$ for the HS chain
with $N=26$ and $m=2$.
\label{fig:diN26m2}}
\end{figure}

We shall next discuss in more detail to what extent the first condition fails. To this end,
we have used the formula for the partition function of the HS chain given in~\cite[eq.~(22)]{FG05}
to compute the spectrum for a wide range of values of $m$ and $N$ such that $m^N\le 2^{28}$. Our results evidence that when
$N$ is sufficiently large the vast majority of the differences $d_i\equiv\cE_{i+1}-\cE_i$
are equal to $d=1$ (for even $N$) or $d=2$ (for odd $N$), as shown in fig.~\ref{fig:percent}.
Thus, we can say that the spectrum of the HS chain is {\em quasi}-equispaced for large $N$.
Moreover, our computations indicate that the differences $d_i\ne d$
correspond to energies $\cE_i$ in the tail of the Gaussian distribution $g(\cE)$; see, \emph{e.g.},
fig.~\ref{fig:diN26m2} for the case $m=2$, $N=26$.

\begin{figure}[t]
\centering
\includegraphics[width=8cm]{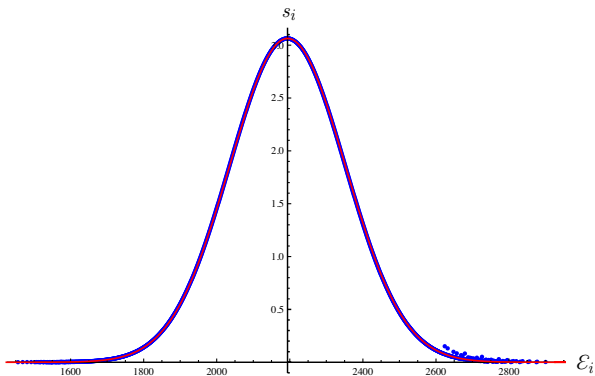}
\caption{Comparison of the normalized spacings $s_i$ (blue dots) with
the Gaussian distribution~\eqref{sid} (continuous red curve) for the HS chain
with $N=26$ and $m=2$.
\label{fig:si}}
\end{figure}

The two features of the spectrum of the HS chain just described provide the key to understanding why
the formula~\eqref{P}, with $\smax$ given by~\eqref{smax}, still works remarkably well in this case. Indeed,
since the level density is approximately Gaussian, we have
$$
\eta(\cE)=\int_{-\infty}^\cE g(t)\ms\diff t\,,
$$
so that
$$
\eta_{i+1}-\eta_i\equiv\eta(\cE_{i+1})-\eta(\cE_i)\simeq\eta'(\cE_i)d_i=g(\cE_i)d_i\,.
$$
By condition (iii) in section~1 we have $\eta_n\simeq 1$ and $\eta_1\simeq 0$, and thus
$\De\simeq1/(n-1)$. Hence, the normalized spacings $s_i$ are approximately given by
$$
s_i=\frac{\eta_{i+1}-\eta_i}\De\simeq(n-1)d_i\ms g(\cE_i)\equiv
\frac{(n-1)d_i}{\sqrt{2\pi}\si}\,\e^{-\frac{(\cE_i-\mu)^2}{2\si^2}}\,.
$$
Since the few values of $d_i$ different from $d$ correspond to energies $\cE_i$
several standard deviations away from $\mu$, their associated spacings $s_i$
turn out to be very small. Therefore, we can write with great accuracy
\begin{equation}\label{sid}
s_i\simeq\frac{(n-1)d}{\sqrt{2\pi}\si}\,\e^{-\frac{(\cE_i-\mu)^2}{2\si^2}}
\equiv\smax\,\e^{-\frac{(\cE_i-\mu)^2}{2\si^2}}\,,
\end{equation}
except for a few spacings $s_i$ very small compared to $\smax$; see, \emph{e.g.},
fig.~\ref{fig:si} for the case $N=26$, $m=2$. As shown in ref.~\cite{BFGR08},
 eq.~\eqref{sid} leads directly to the approximate formula~\eqref{P} for the
 cumulative spacings distribution $P(s)$. This explains why the latter formula
 is also valid for the HS chain, with the value of $\smax$ given in eq.~\eqref{smax}.

\acknowledgments
This work was partially supported by the DGI under grant no.~FIS2005-00752, and
by the Complutense University and the DGUI under grant no.~GR74/07-910556.
J.C.B. acknowledges the financial support of the Spanish Ministry of Education
and Science through an FPU scholarship.


\end{document}